\documentclass[12pt]{amsart}

\usepackage{amsmath,amssymb,amsthm,a4,bm,epsfig,color}

\newtheorem{theorem}{Theorem}
\newtheorem{proposition}{Proposition}
\newtheorem{lemma}{Lemma}
\newtheorem{remark}{Remark}

\newcommand{\bbR}{{\mathbb R}}
\newcommand{\bbC}{{\mathbb C}}
\newcommand{\bbN}{{\mathbb N}}
\newcommand{\Hone}{H^{1}({\mathbb R}^N,{\mathbb C})}

\newcommand{\cL}{{\mathcal L}}
\newcommand{\cM}{{\mathcal M}}

\newcommand{\cE}{{\mathcal E}}
\newcommand{\cT}{{\mathcal T}}
\newcommand{\cR}{{\mathcal R}}

\newcommand{\esssup}{ess\,sup}
\renewcommand{\Re}{{\mathrm{Re}}}
\renewcommand{\Im}{{\mathrm{Im}}}

\begin{document}

\title[]{Colliding solitons for the nonlinear Schr\"odinger equation}

\author[]{W. K. Abou Salem $^{1, \#}$, J. Fr\"ohlich $^{2}$ and I. M. Sigal $^{1, \#}$}

\address{$^1$ Department of Mathematics, University of Toronto, Toronto, Ontario, Canada M5S 2E4. E-mail: walid@math.utoronto.ca;  im.sigal@utoronto.ca.} 
\address{$^2$ Institute for Theoretical Physics, ETH Zurich, Zurich CH-8093, Switzerland. E-mail: juerg@itp.phys.ethz.ch; Institut des Hautes \'Etudes Scientifique, F-91440 Bures-sur-Yvette, France. E-mail:juerg@ihes.fr.}
\address{$^\#$ Supported in part by NSERC grant NA 7901. }
\maketitle

\begin{abstract}
We study the collision of two fast solitons for the nonlinear Schr\"odinger equation in the presence of a spatially adiabatic external potential. For a high initial relative speed $\|v\|$ of the solitons, we show that, up to times of order $\log\|v\|$ after the collision, the solitons preserve their shape (in $L^2$-norm), and the dynamics of the centers of mass of the solitons is approximately determined by the external potential, plus error terms due to radiation damping and the extended nature of the solitons. We remark on how to obtain longer time scales under stronger assumptions on the initial condition and the external potential.
\end{abstract}


\section{Introduction}\label{sec:Introduction}

In this paper, we study the collision of two fast solitons in the presence of a (time-dependent) external potential that varies slowly in space compared to the {\it size} of the solitons. We show, for a class of typical local and nonlocal nonlinearities, that if the initial relative speed of the solitons is $\|v\|\gg 1$ and the spatial variation of the external potential is sufficiently slow, then the solitons pass each other almost blindly: The $L^2$-norm of the  difference between the true solution and the one corresponding to a configuration of two solitons moving in the external potential decays algebraically with $\|v\|,$ up to times of order $\log\|v\|,$ after the collision.  This is an example where the solitary waves for NLS display both their {\it ``wave''} and {\it ``particle''} nature. They pass each other almost blindly because they are localized waves with high relative speed and relative phase, while their center of mass dynamics is approximately that of a classical particle in a spatially adiabatic external potential.

The problem of asymptotic behaviour of multi-soliton configurations (scattering theory) for the nonlinear Schr\"odinger equation {\it without} an external potential has been addressed in \cite{P} and \cite{RSS}; see also \cite{MMT}. In these papers, the authors prove, under rather strong spectral assumptions on the linearized equation, the asymptotic stability of multi-solitons in three (or higher) dimensions. The main ingredient of their analysis is asymptotic stability of single solitons and dispersive estimates (which are related to the ``charge-transfer model''). Here, our results and approach are different: We study the long-time dynamics of the collision of fast solitons in the presence of an external potential rather than the asymptotics, and we use softer yet more robust techniques that allow for treating a wide class of systems under weak assumptions. Furthermore, our analysis holds in any dimension $N\ge 1.$

There has been considerable progress in understanding the long-time dynamics of single solitons in spatially adiabatic external potentials and in the presence of nonlinear perturbations, \cite{FJGS1,FJGS2,HZ1,A-S1,A-S2, HZ2}. The analysis below together with additional mild spectral assumptions can be extended to study the effective dynamics of multiple solitons with low velocities in slowly varying external potentials (and in the presence of nonlinear perturbations) as long as the soliton centers of mass are {\it well separated}. \footnote{We note that for the generalized KdV equation, there has been some recent progress in understanding the collision of a fast thin soliton with a slow broad soliton in the absence of an external potential, see \cite{MM1,MM2}; and also \cite{T} for a recent review about problems related to the stability of solitons.}

\subsection{Description of the problem}

We consider the nonlinear Schr\"odinger equation
\begin{equation}
i\partial_t \psi(x,t) = (-\Delta + V_h(x,t)) \psi(x,t) - f(\psi(x,t)), \label{eq:NLSE}
\end{equation}
where $\Delta = \sum_{i=1}^N\frac{\partial^2}{\partial x_i^2}$ is the $N$-dimensional Laplacian, with $N\ge 1,$ $V_h$ denotes the (time-dependent) external potential, with $$V_h(x,t)\equiv V(hx,t),$$ and $f$ is a focusing nonlinearity 
\begin{equation*}
f: H^1(\bbR^N ; \bbC) \rightarrow H^{-1}(\bbR^N ; \bbC),
\end{equation*}
such that $\overline{f(\psi)}=f(\overline{\psi}).$

We now discuss the various assumptions we make, which are simultaneously satisfied by typical local and Hartree nonlinearities, see Remark \ref{rem:nonlinearity} below.
 
\begin{itemize}
\item[(A1)] {\it Global well-posedness.} The nonlinear Schr\"odinger equation (\ref{eq:NLSE}) is globally well-posed in $H^1.$ 
\end{itemize}
We refer the reader to \cite{Ca}, chapter 6, for well-posedness of (\ref{eq:NLSE}) in energy space for time-independent potentials, and \cite{A-S1} for the case of time-dependent external potentials and nonlinearities. We make the following assumption on the regularity and symmetries of the nonlinearity.

\begin{itemize}
\item[(A2)] {\it Nonlinearity.} Let $F: H^1 \rightarrow \bbR$ be the functional such that its Fr\'echet derivative $F'=f.$ We assume that $F\in C^3(H^1 ; \bbR)$ and that $F(T\cdot)=F(\cdot),$ where $T$ is a translation 
\begin{equation*}
T_a^{tr}: u(x)\rightarrow u(x-a), \ \ a\in \bbR^N,
\end{equation*}
a rotation 
\begin{equation*}
T_R^r: u(x)\rightarrow u(R^{-1}x), \ \ R\in SO(N),
\end{equation*}
a gauge transformation
\begin{equation*}
T_\gamma^g : u(x)\rightarrow e^{i\gamma}u(x), \ \ \gamma \in [0,2\pi),
\end{equation*}
or a boost 
\begin{equation*}
T_v^b: u(x)\rightarrow e^{\frac{i}{2}v\cdot x} u(x), \ \ v\in \bbR^N.
\end{equation*}
\end{itemize}

We are interested in the dynamics of multi-solitons, so we assume the existence of solitary wave solutions when $V=0;$ see for example \cite{Ca}, chapter 8, for a discussion of solitary waves for NLS.
\begin{itemize}
\item[(A3)] {\it Solitary waves.} When $V=0,$ there exists an  interval $I\subset \bbR$ such that, for all $\mu\in I,$ (\ref{eq:NLSE}) admits solitary wave solutions of the form
\begin{equation*}
u_\sigma = e^{i\mu t + i\gamma +  \frac{i}{2}v\cdot (x-a-vt)}\eta_\mu(x-a-vt),
\end{equation*}
where $$\sigma = (a,v,\gamma,\mu)\in \bbR^N \times \bbR^N \times [0,2\pi) \times I.$$
Here, $\eta_\mu$ is a positive and spherically symmetric function satisfying the nonlinear eigenvalue problem 
\begin{equation}
\label{eq:NLEV}
(-\Delta +\mu)\eta_\mu -f(\eta_\mu)=0,
\end{equation}
$$\eta_\mu\in L^2(\bbR^N)\cap C^2(\bbR^N) ,$$ 
\begin{equation}
\label{eq:SolDecay}
\||x|^3\eta_\mu\|_{L^2}+\||x|^2|\nabla \eta_\mu|\|_{L^2} + \||x|^2\partial_\mu \eta_\mu\|_{L^2}<\infty, \forall \mu \in I,
\end{equation}
and 
$$\eta_\mu \propto e^{-\sqrt{\mu}\|x\|} \; \mathrm{as} \; \|x\|\rightarrow\infty.$$  Let 
$$m(\mu)=\frac{1}{2}\int dx ~\eta_{\mu}^2,$$ the ``charge'' of the soliton. We assume that $$\partial_\mu m(\mu)>0,$$ which implies {\it orbital stability}, see \cite{GSS1,GSS2,SS}.

\end{itemize}

We require some local properties of the nonlinearity, which are satisfied for classes of local and nonlocal nonlinearities, see Remark \ref{rem:nonlinearity}.

\begin{itemize}

\item[(A4)] {\it Localization}.
We assume that 
\begin{equation*}
\|(f'(\eta_{\mu_1}+u_{\sigma_2}) - f'(\eta_{\mu_1})) X \eta_{\mu_1}\|_{L^2} \le C e^{-\xi\|a_2\|},
\end{equation*}
and
\begin{equation*}
\|(f(\eta_{\mu_1}+u_{\sigma_2})-f(\eta_{\mu_1}))X \eta_{\mu_1}\|_{L^1}\le C e^{-\xi\|a_2\|}
\end{equation*}
where $\eta_{\mu_1} \equiv u_{(0,0,0,\mu_1)},$ $X=1,\, x$ or $\partial_x.$
Here, $\xi\in (0,\min(\sqrt{\mu_1},\sqrt{\mu_2}))$ and $C>0$ are constants that are independent of $a_2$ and $v_2.$

\item[(A5)]  For $g\in L^p({\mathbb R}^N), \; p\ge1,$  $u= \sum_{i=1}^n u_{\sigma_i},$ where $u_\sigma$ appears in (A3), and $w\in H^1$ with $\|w\|_{L^2}\le 1,$ we have 
\begin{equation*}
|\langle g, f(u+w) - f(u) - f'(u)w\rangle |\le C \|w\|^2_{L^2},
\end{equation*}
where $C$ is a constant the depends on $g$ and $\mu_i, i=1,\cdots ,n.$
\end{itemize}

We make the following assumption on the external potential, which, among other things, guarantees well-posedness of (\ref{eq:NLSE}) in $H^1,$ inspite of the fact that the energy in no more conserved, see \cite{A-S1}.

\begin{itemize}
\item[(A6)] The external potential $V\in W^{1,\infty}(\bbR; C^2(\bbR^N)).$
\end{itemize}

We now discuss the initial condition.We are interested in the collision of solitons with high relative speed. A 2-soliton configuration plus a fluctuation is given by 
\begin{equation}
\psi(t=0)=\phi (x) = e^{\frac{i}{2}\widetilde{v}_{1}\cdot x}\eta_{\widetilde{\mu}_1}(x-\widetilde{a}_{1}) + e^{\frac{i}{2}\widetilde{v}_{2}\cdot x}\eta_{\widetilde{\mu}_2}(x-\widetilde{a}_{2}) + \widetilde{w} , \label{eq:InitialCond}
\end{equation}
with $\widetilde{a}_1,\widetilde{a}_2,\widetilde{v}_1,\widetilde{v}_2\in \bbR^N,$ $\widetilde{w} \in H^1,$ and $\widetilde{\mu}_1,\widetilde{\mu}_2 \in I_0,$ where $I_0\subset I\backslash \partial I$ is a bounded interval such that its closure $\overline{I}_0\subset I\backslash \partial I.$ We assume that 
$$\|\widetilde{v}_1-\widetilde{v}_2\|\gg (\inf_{\mu\in I_0}m'(\mu))^{-1}$$ with
$\frac{\|\widetilde{a}_1-\widetilde{a}_2\|}{\|\widetilde{v}_1-\widetilde{v}_2\|}=O(1).$  

We assume that the fluctuation $\widetilde{w} $ is {\it small}. More specifically, $\widetilde{w} \in H^1$ such that $$\|\widetilde{w} \|^2_{L^2} < C \|\widetilde{v}_1-\widetilde{v}_2\|^{-1}.$$ 

We did not impose any condition on the directions of the relative speed and position of the solitons. In particular, we can have 
$$(\widetilde{a}_1-\widetilde{a}_2)\cdot (\widetilde{v}_1-\widetilde{v}_2) <0,$$
which is the case corresponding to {\it colliding} solitons. We remark later how one obtains better estimates in case the solitons are {\it escaping} each other.

In what follows, we denote by $v_0:= \widetilde{v}_1-\widetilde{v}_2,$ the initial relative velocity of the solitons. 

\subsection{Main result}\label{sec:Main}

We are in a position to state our main result, whose generalization for fast $n$-solitons, $n\ge 2,$ is straight forward.

\begin{theorem}\label{th:Main}
Consider the nonlinear Schr\"odinger equation (\ref{eq:NLSE}) with initial condition given by (\ref{eq:InitialCond}), and suppose assumptions (A1)-(A6) hold. Then, for any fixed $\alpha\in (0,1),$ $\|v_0\|\gg (\inf_{\mu\in I_0} m'(\mu))^{-\frac{2}{1-\alpha}}$ and $h\ll (\inf_{\mu\in I_0} m'(\mu))^{\frac{1}{1-\alpha}},$  the solution of the initial value problem can be written as 
\begin{equation*}
\psi (x,t) = e^{i\gamma_1 t + \frac{i}{2} v_1\cdot(x-a_1)} \eta_{\mu_1}(x-a_1) + e^{i\gamma_2 t + \frac{i}{2}v_2\cdot (x-a_2)}\eta_{\mu_2}(x-a_2) + w(x,t),
\end{equation*}
for all $t\in [0, \tau_{  \alpha  }), \, \tau_  \alpha  := C   \alpha   \min(\log\|v_0\|,2|\log h|),$ and 
\begin{equation}
\label{eq:SupFluctuation}
\sup_{t\in [0, \tau_  \alpha  ) }\|w\|_{L^2} \le C'(\|v_0\|^{-\frac{1-  \alpha  }{2}}+ h^{1-  \alpha  }),
\end{equation}
where the constants $C,C'>0$ are independent of $v_0, h$ and $\alpha .$
Furthermore, the parameters $a_i,v_i,\gamma_i, \mu_i, i=1,2,$ satisfy, for $t\in [0,\tau_  \alpha  ),$ the following equations
\begin{align*}
&\partial_t a_i = v_i + O(\|v_0\|^{-(1-  \alpha)}+ h^{2(1-  \alpha)}+e^{-\xi \|a_1-a_2\|}) , \\
&\partial_t v_i = - 2\nabla V_h(a_i,t) +O(\|v_0\|^{-(1-  \alpha)}+ h^{2(1-  \alpha)}+e^{-\xi \|a_1-a_2\|}) ,\\
&\partial_t \gamma_i = \mu_i + \frac{v_i^2}{4} - V_h(a_i,t) + O(\|v_0\|^{-(1-  \alpha)}+ h^{2(1-  \alpha)}+e^{-\xi \|a_1-a_2\|}) ,\\
&\partial_t \mu_i =  O(\|v_0\|^{-(1-  \alpha)}+ h^{2(1-  \alpha)}+e^{-\xi \|a_1-a_2\|}),
\end{align*}
for some $\xi\in(0,\min(\sqrt{\mu_1},\sqrt{\mu_2}))$ that is independent of $\|v_0\|$ and $h.$
\end{theorem}

In particular, for $\|v_0\|\gg 1$ sufficiently large, and $h = O(\|v_0\|^{-\frac{1}{2}}),$ the solitons preserve their shape, in $L^2$-norm, up to times $\log\|v_0\|$ after the collision, such that the dynamics of the centers of mass of the solitons is approximately determined by the Hamilton equations for two classical particles in the external potential.

Our analysis relies on three main ingredients. First, using a skew-orthogonal (or Lyapunov-Schmidt) decomposition property (Proposition \ref{pr:SOD}, Sect. \ref{sec:SOD}), we decompose the solution of (\ref{eq:NLSE}) with initial condition close to a 2-soliton configuration, as described by (\ref{eq:InitialCond}), into a path belonging to a symplectic manifold of 2-soliton states, and a part describing a fluctuation skew-orthogonal to the manifold. The dynamics on the 2-soliton manifold is obtained by the skew-orthogonal projection of the Hamiltonian flow generated by the nonlinear Schr\"odinger equation in a small tubular neighbourhood of the 2-soliton manifold onto the latter (Proposition \ref{pr:RepEqMot}, Sect. \ref{sec:RepEqMot}). As for the  fluctuation, we control its $L^2$-norm using charge conservation and skew-orthogonal decomposition (Proposition \ref{pr:ControlFluctuation}, Sect. \ref{sec:ControlFluctuation}). The main difference between our approach and the one for studying the effective dynamics of a single soliton in an external potential, as for example in  \cite{FJGS1}, is that we control the $L^2$-norm of the fluctuation using charge conservation, rather than controlling its $H^1$-norm by using an approximate Lyapunov functional and proving constraint positivity of the {\it Hessian}, Eq. (\ref{eq:Hessian}) below, under additional assumptions that are verified in the case of special local nonlinearities. Unlike the $L^2$-norm, the $H^1$-norm of $\psi$ grows like $\|v_0\|,$ and we lose control over $\|w\|_{H^1}$ as $\|v_0\|\rightarrow\infty.$


\begin{remark}\label{rem:nonlinearity}
We now give some concrete examples for which assumptions (A1)-(A5) are simultaneously satisfied.

An example where assumptions (A1) - (A3) are satisfied is when $f$ is a Hartree nonlinearity, $$f(\psi) =  (W\star |\psi|^2)\psi,$$
such that $W$ is positive, spherically symmetric, belongs to  $L^p+L^\infty,$ with $p>\frac{N}{2}, p\ge 1,$ and decays at infinity, $W\rightarrow 0 $ as $\|x\|\rightarrow\infty;$ see \cite{Ca,A-S1}.   The localization property, assumption (A4), is satisfied if in addition $W$ decays exponentially fast. We now verify that (A5) holds for $p\ge 2.$ From the form of the nonlinearity, we have 
\begin{equation*}
|\langle g, f(u+w) - f(u) - f'(u)w\rangle | \le C (|\langle g,( W\star |w|^2) u\rangle| + |\langle g,( W\star |\overline{u}w| ) w\rangle|).
\end{equation*}
Applying H\"older's and Young's inequalities, we have 
\begin{align*}
|\langle g, (W\star |w|^2) u\rangle| &\le \|g u\|_{L^{q'}}\|W\star |w|^2\|_{L^q} \\
&\le \|g\|_{L^{q'}}\|u\|_{L^\infty} \|W\|_{L^q} \||w|^2\|_{L^1}\\
&\le \|g\|_{L^{q'}}(\sum_{i=1}^n\|u_i\|_{L^\infty}) \|W\|_{L^q} \|w\|_{L^2}^2\\
&\le C \|w\|_{L^2}^2,
\end{align*}
where $q=p$ or $\infty$ ($W\in L^p+L^\infty$) and $1=1/q + 1/q'.$ Similarly, 
\begin{align*}
 |\langle g,(W\star |\overline{u}w| ) w\rangle| &\le \|g W\star |\overline{u}w| \|_{L^2}\|w\|_{L^2} \\
 &\le \|g\|_{L^{\frac{2q}{q-2}}}\|W\star (|u||w|)\|_{L^q} \|w\|_{L^2} \\
 &\le \|g\|_{L^{\frac{2q}{q-2}}}\|W\|_{L^q}\|uw\|_{L^1} \|w\|_{L^2}\\
  &\le \|g\|_{L^{\frac{2q}{q-2}}}\|W\|_{L^q} \|u\|_{L^2}\|w\|_{L^2}^2 \\
 &\le \|g\|_{L^{\frac{2q}{q-2}}}\|W\|_{L^q} (\sum_{i=1}^n\|\eta_{\mu_i}\|_{L^2})\|w\|_{L^2}^2 \\
 &\le C \|w\|_{L^2}^2.
\end{align*}
Therefore, assumption (A5) is satisfied.

Another example where the various assumptions are satisfied is when $f$ is a local nonlinearity. For example, (A1) and (A2) are satisfied if $f$ is of the form 
\begin{equation*}
f(\psi)(x) = h(|\psi(x)|^2) \psi(x),
\end{equation*}
where $h\in C^2({\mathbb R}^+,{\mathbb R})$ with 
$$\partial_r^{k}h(r) \le C (1+r^{\alpha - k}), \; k=0,1,2,$$
$\alpha\in (0,\frac{2}{N-2}),\; N\ge 3,$ and  $\alpha\in (0,\infty)$ if $N=1,2;$ see for example \cite{Ca,A-S1} for a discussion of well-posedness in $H^1.$ Solitary wave solutions appearing in (A3) exist, if, in addition, 
\begin{align*}
&-\infty < \lim_{r\rightarrow \infty} h(r) < \mu\\
&-\infty \le \lim_{r\rightarrow\infty} r^{-\alpha} h(r) \le C,
\end{align*}
and there exists $r_0 >0$ with 
$$\int_0^{r_0} h(r)dr>\mu r_0,$$ 
see \cite{BL1,BL2}. The condition of orbital stability can to be checked for each nonlinearity, see \cite{GSS1,GSS2,SS}. Assumption (A4) follows directly from (A3) and the form of the local nonlinearity. Furthermore, assumption (A5) is satisfied if 
\begin{equation*}
\sup_{r\in {\mathbb R}^+} r^{\frac{2k-1}{2}}\partial_r^{k} h(r) <\infty ,\; k=1,2.
\end{equation*}
An explicit example of a local nonlinearity that satisfies all the above hypotheses is 
$$f(\psi) = |\psi|^s \psi \, \chi_{\theta,s}(|\psi|) ,\; s\in (0,\frac{4}{N}),$$
where $\chi_{\theta,s}, \; \theta \gg 1,$ is a smooth regularization which is chosen such that (A5) is satisfied. For example, 
\begin{equation*}
\chi_{\theta,s}(y) = 
\begin{cases}
1 , \;\mathrm{if} \; |y|^{sgn(s-1)} < \theta/2 \\
|y|^{1-s},\;\mathrm{if} \; |y|^{sgn(s-1)} > \theta 
\end{cases}.
\end{equation*}


More generally, $f$ can be a sum of both local and nonlocal nonlinearities. 
\end{remark}

\begin{remark}\label{rem:LongerTime}
We now remark on special cases where one can obtain a control of the fluctuation over different (and longer) time scales. Assume (A1)-(A6) hold, and suppose, for the sake of simplicity, that $h=0,$ which corresponds to a spatially flat potential.

\begin{itemize}

\item[(1)]Large separation. If the soliton centers of mass are initially separated by a distance $d \gg \max(\frac{1}{\sqrt{\mu_1}},\frac{1}{\sqrt{\mu_2}},|\log \inf_{\mu\in I_0} m'(\mu)|)$ and $\|\widetilde{v}_1\|,\|\widetilde{v}_2\|=O(1),$ with $\|\widetilde{w} \|_{L^2} = O(e^{-\chi d})$ for some $\chi>0,$ then one obtains a result similar to Theorem \ref{th:Main} such that $\sup_{t\in [0, d^\epsilon)}\|w\|_{L^2}<C/\sqrt{d},$ for any $\epsilon\in (0,1)$ and, for $t\in [0, d^\epsilon),$ 
\begin{align*}
&a_i(t) =  \widetilde{a}_i + t\widetilde{v}_i + O(d^{-(1-\epsilon)})\\
&v_i(t) = \widetilde{v}_i + O(d^{-(1-\epsilon)})\\
&\mu_i(t) = \widetilde{\mu}_i + O(d^{-(1-\epsilon)})\\
&\gamma_i(t) = \widetilde{\gamma}_i + t(\widetilde{\mu}_i + \frac{\widetilde{v}_i^2}{4} - V(0,t)) + O(d^{-(1-\epsilon)}),
\end{align*}
with $i=1,2,$ see Sect. \ref{rem:ProofLongerTime}.

\item[(2)] Escaping solitons. Suppose that the solitons escape each other with a high relative speed
$$(\widetilde{a}_1-\widetilde{a}_2)\cdot (\widetilde{v}_1-\widetilde{v}_2) \ge 0,$$
and $\|v_0\| \gg (\inf_{\mu\in I_0}m'(\mu))^{-2},$ $\|\widetilde{w} \|_{L^2} = O(e^{-\chi \|v_0\|})$ for some $\chi>0,$ then, for any fixed $\epsilon\in (0,1),$  there exists a contant $C,$ independent of $\|v_0\|$ and $\epsilon,$ such that
\begin{equation*}
\sup_{t\in [0,\|v_0\|^\epsilon)}\|w(t)\|_{L^2} \le C \|v_0\|^{-\frac{1}{2}}
\end{equation*} 
and, for $t\in[0,\|v_0\|^{\epsilon}),$
\begin{align*}
&a_i(t) =  \widetilde{a}_i + t\widetilde{v}_i + O(\|v_0\|^{-1+\epsilon})\\
&v_i(t) = \widetilde{v}_i + O(\|v_0\|^{-1+\epsilon})\\
&\mu_i(t) = \widetilde{\mu}_i + O(\|v_0\|^{-1+\epsilon})\\
&\gamma_i(t) = \widetilde{\gamma}_i + t(\widetilde{\mu}_i + \frac{\widetilde{v}_i^2}{4} - V(0,t)) + O(\|v_0\|^{-1+\epsilon}),
\end{align*}
with $i=1,2,$ see Sect. \ref{rem:ProofLongerTime}.
\end{itemize}
\end{remark}

The organization of this paper is as follows. In Sections \ref{sec:NLSeq}, we recall some basic properties of the nonlinear Schr\"odinger equation. In Section \ref{sec:SolitonManifold}, we recall the soliton manifold, and we introduce the 2-soliton  (or, more generally, $n$-soliton) manifold. In Section \ref{sec:SOD}, we prove the skew-orthogonal decomposition property for elements of neighbourhoods in $H^1$ that are close in ($L^2$-norm) to a two-soliton manifold, which is a central tool in our analysis. In Section \ref{sec:RepEqMot}, we use the skew-orthogonal property and the nonlinear Schr\"odinger equation (\ref{eq:NLSE}) to find the reparametrized equations of motion corresponding to the parameters on the two-soliton manifold, and in Section \ref{sec:ControlFluctuation}, we control the $L^2$-norm of the fluctuation using charge conservation and the skew-orthogonal decomposition. In Section \ref{sec:Proof} we prove Theorem \ref{th:Main} by combining the results of Propositions \ref{pr:SOD}, \ref{pr:RepEqMot} and \ref{pr:ControlFluctuation}. We finally remark on separating solitons in Section \ref{rem:ProofLongerTime}.

\subsection{Notation}

\begin{itemize}
\item In the following, $L^p(I)$  denotes the standard Lebesgue space, $1\le p\le \infty,$ with norm
\begin{align*}
&\|f\|_{L^p} = (\int_I dx~ |f(x)|^p)^{\frac{1}{p}}, \ \ f\in L^p(I),  p<\infty , \\ &\|f\|_{L^\infty} = \esssup(|f|) , \ \ f\in L^\infty (I).
\end{align*}

\item We denote by $\langle \cdot ,\cdot \rangle $ the scalar product in $L^2(\bbR^N),$ 
\begin{equation*}
\langle u, v\rangle = \Re\int_{\bbR^N} u\overline{v}, \ \ u,v\in L^2(\bbR^N).
\end{equation*}

\item Given the multi-index $\overline{\alpha}= (\alpha_1,\cdots ,\alpha_N) \in \bbN^N,$ we denote $|\overline{\alpha}| = \sum_{i=1}^N \alpha_i.$ Furthermore, $\partial_x^{\overline{\alpha}} := \partial_{x_1}^{\alpha_1}\cdots \partial_{x_N}^{\alpha_N}.$
\item For $1\le p\le\infty$ and $s\in\bbN,$ the (complex) Sobolev space is given by
\begin{equation*}
W^{s,p}(\bbR^N) := \{u\in {\mathcal S}'(\bbR^N), \; \partial_x^{\overline{\alpha}} u\in L^p(\bbR^N), |\overline{\alpha}|\le s\}, 
\end{equation*}
where ${\mathcal S}'(\bbR^N)$ is the space of tempered distributions. We equip $W^{s,p}$ with the norm 
\begin{equation*}
\|u\|_{W^{s,p}} = \sum_{\alpha, |\mathbf{\alpha}|\le s} \|\partial_x^{\overline{\alpha}} u \|_{L^p},
\end{equation*}
which makes it a Banach space. We use the shorthand $W^{s,2}=H^s.$

\item Given $f$ and $g$ real functions on $\bbR^N,$ we denote their convolution by $\star,$
\begin{equation*}
f\star g(x):= \int dy~ f(y) g(x-y) .
\end{equation*}
\end{itemize}

\subsection{Acknowledgements} W.A.S. thanks Catherine Sulem for pointing out references \cite{MM1,MM2,T}.


\section{Hamiltonian structure of the nonlinear Schr\"odinger equation}\label{sec:NLSeq}

In this section, we recall some basic properties of the nonlinear Schr\"odinger equation (\ref{eq:NLSE}), see for example \cite{SS,FJGS1}. We will use these properties in the following sections. 

The space $\Hone$ has a real inner product (Riemannian metric) 
\begin{equation}
\langle u,v \rangle :=\Re \int dx ~ u\overline{v}  
\end{equation}
for $u,v\in H^1(\bbR^N,\bbC).$ \footnote{The tangent space at $\psi\in H^1$ is $\cT_\psi H^1=H^1.$} It is equipped with a symplectic ``form'' 
\begin{equation}
\omega(u,v):= \Im \int  dx ~ u\overline{v} = \langle u,iv\rangle.
\end{equation}
The Hamiltonian functional corresponding to the nonlinear Schr\"odinger equation (\ref{eq:NLSE}) is
\begin{equation}
\label{eq:NLHamiltonian}
H_V(\psi):=\frac{1}{2}\int |\nabla\psi|^2 dx +\frac{1}{2}\int V |\psi|^2 - F(\psi).
\end{equation}
Using the correspondence 
\begin{align*}
\Hone &\longleftrightarrow H^1(\bbR^N,\bbR) \oplus H^1(\bbR^N,\bbR)\\
\psi &\longleftrightarrow (\Re\psi \; \Im\psi)\\
i^{-1}&\longleftrightarrow J,
\end{align*}
where $J:=\begin{pmatrix} 0 & 1 \\ -1 & 0 \end{pmatrix}$ is the complex structure on $H^1(\bbR^N,\bbR^{2}),$ the nonlinear Schr\"odinger equation can be written as 
\begin{equation*}
\partial_t \psi = J H_V' (\psi).
\end{equation*}
Furthermore, 
\begin{align*}
\langle u,v\rangle &= \int dx ~ (\Re u \; \Im u) \left (
\begin{matrix}
\Re v \\ \Im v
\end{matrix}
\right ),\\
\omega (u,v) &= \int dx~  (\Re u \; \Im u) \left (
\begin{matrix}
0 & -1 \\
1 & 0
\end{matrix}
\right )
\left (
\begin{matrix}
\Re v \\ \Im v
\end{matrix}
\right ).
\end{align*}

We note that since the Hamiltonian functional $H_V$ defined in (\ref{eq:NLHamiltonian}) is nonautonomous, the energy is {\it not } conserved. For $\psi\in H^1$ satisfying (\ref{eq:NLSE}), $$\partial_t H_V(\psi)= \frac{1}{2}\int dx ~ (\partial_tV) |\psi|^2,$$ see \cite{A-S1} for a proof of this statement. Still, $H_V$ is invariant under global gauge transformations, 
\begin{equation*}
H_V(e^{i\gamma}\psi)=H_V(\psi),
\end{equation*}
and the associated conserved quantity is the ``charge'' 
\begin{equation*}
N(\psi):= \frac{1}{2}\int dx~ |\psi|^2.
\end{equation*}

The assumption $\partial_\mu m(\mu)>0$ implies that $\eta_\mu$ appearing in assumption (A3) is a local minimizer of $H_{V=0}(\psi)$ restricted to the balls ${\mathcal B}_m:= \{\psi\in H^1 : N(\psi)=m\},$ for $m>0;$ see \cite{GSS1,GSS2}. They are critical points of the functional 
\begin{equation}
\label{eq:EnergyFunctional}
\cE_\mu (\psi) := \frac{1}{2} \int dx~ (|\nabla\psi|^2 +\mu|\psi|^2)-F(\psi),
\end{equation}
where $\mu=\mu(m)$ is a Lagrange multiplier.

\section{Soliton Manifolds}\label{sec:SolitonManifold}
In this section, we recall the definition and properties of a single soliton manifold (see \cite{FJGS1,FJGS2,HZ1,A-S1,A-S2,HZ2}), and we introduce the multi-soliton manifold.

\subsection{Soliton Manifold}
We introduce the combined transformation $T_{av\gamma},$ which is given by
\begin{equation*}
\psi_{av\gamma}:= T_{av\gamma}\psi = e^{i(\frac{1}{2}v\cdot (x-a)+\gamma)}\psi(x-a),
\end{equation*}
where $v,a\in\bbR^N$ and $\gamma\in [0,2\pi).$ We define the soliton manifold as 
\begin{equation*}
\cM_s:= \{\eta_\sigma:= T_{av\gamma}\eta_\mu , \; \sigma= (a,v,\gamma,\mu) \in \bbR^N\times\bbR^N\times [0,2\pi)\times I \},
\end{equation*}
where $I$ appears in assumption (A3). If $f'(0)=0,$ where $f$ appears in (\ref{eq:NLSE}), then $I\subset \bbR^+.$ 

The tangent space to the soliton manifold $\cM_s$ at $\eta_\mu\in \cM_s$ is given by 
\begin{equation*}
\cT_{\eta_\mu}\cM_s = span\{ E_t,E_g,E_b,E_s\},
\end{equation*}
where
\begin{align*}
E_t &:= \nabla_a T_a^{tr}\eta_\mu|_{a=0}=-\nabla\eta_\mu\\
E_g &:= \partial_\gamma T_\gamma^g \eta_\mu|_{\gamma=0} = i\eta_\mu\\
E_b &:= 2\nabla_v T_v^{b}\eta_\mu|_{v=0}=ix\eta_\mu\\
E_s &:= \partial_\mu\eta_\mu.
\end{align*}
In the following, we denote by 
\begin{align}
&e_j   := -\partial_{ x_j  }, \;  j=1,\cdots ,N,\nonumber \\ 
&e_{j+N} := i  x_j  , \; j=1,\cdots ,N, \nonumber\\
&e_{2N+1} := i, \nonumber\\
&e_{2N+2} := \partial_\mu, \label{eq:TangentBases}
\end{align}
which, when acting on $\eta_\sigma\in\cM_s,$ generate the basis vectors $\{e_  \alpha   \eta_\sigma\}_{  \alpha  =1}^{2N+2}$ of $\cT_{\eta_\sigma}\cM_s.$

The soliton manifold $\cM_s$ inherits a symplectic structure from $(H^1,\omega).$ For $\sigma = (a,v,\gamma,\mu) \in \bbR^N\times\bbR^N\times [0,2\pi)\times I,$
\begin{equation*}
\Omega_\sigma :=  P_\sigma J^{-1} P_\sigma ,
\end{equation*}
where $P_\sigma$ is the $L^2$-orthogonal projection onto $\cT_{\eta_\sigma}\cM_s.$ 

We have the following easy lemma, which we prove in the Appendix.

\begin{lemma}\label{lm:Metric}
If $\partial_\mu m(\mu)>0,$ then $\Omega_\sigma$ is invertible.
\end{lemma}
Explicitly, we have
\begin{align}
\Omega_{\sigma}|_{\cT_{\eta_\sigma}\cM_s} :&= \{\langle e_  \alpha  \eta_\sigma, i e_\beta   \eta_\sigma \rangle\}_{1\le \alpha,\beta\le 2N+2}\nonumber \\
&= 
\left(
\begin{matrix}
0 & -m(\mu) {\mathbf 1}_{N\times N} & 0 & -\frac{1}{2}vm'(\mu) \\
m(\mu) {\mathbf 1}_{N\times N} & 0 & 0 & am'(\mu) \\
0 & 0 & 0 & m'(\mu) \\
\frac{1}{2}v^T m'(\mu) & -a^Tm'(\mu) & -m'(\mu) & 0
\end{matrix}
\right),
\label{eq:Metric}
\end{align}
where ${\mathbf 1}_{N\times N}$ is the $N\times N$ identity matrix, and $(\cdot)^T$ stands for the transpose of a vector in $\bbR^N;$ see the proof of Lemma \ref{lm:Metric} in the Appendix. 

\subsection{Group structure}
The anti-selfadjoint operators $\{e_  \alpha  \}_{\alpha=1,\cdots,2N+1}$ defined in (\ref{eq:TangentBases}) form the generators of the Lie algebra ${\mathsf g}$ corresponding to the Heisenberg group ${\mathsf H}^{2N+1},$ where the latter is given by 
\begin{equation*}
(a,v,\gamma)\cdot(a',v',\gamma')=(a'',v'',\gamma''),
\end{equation*}
with $a''=a+a',$ $v''=v+v',$ and $\gamma''=\gamma'+\gamma+\frac{1}{2}v\cdot a'.$\footnote{This structure was noted for the case $N=1$ in \cite{HZ1}.}
Elements of ${\mathsf g}$ satisfy the commutation relations
\begin{equation}
[e_i,e_{j+N}]=-e_{2N+1}\delta_{ij}, \ \ i,j= 1,\cdots,N , \label{eq:CommutationRelations}
\end{equation}
and the rest of the commutators are zero.

\subsection{Zero modes}
The solitary wave solutions transform covariantly under translations and gauge transformations, i.e.,
\begin{equation*}
{\mathcal E}_\mu' (T_a^{tr} T_\gamma^g \eta_\mu)=0
\end{equation*}
for all $a\in \bbR$ and $\gamma\in [0,2\pi).$ There are zero modes of the {\it Hessian},   
\begin{equation}
\label{eq:Hessian}
\cL_\mu:= -\Delta +\mu -f'(\eta_\mu),
\end{equation} 
associated to these symmetries. We have the following lemma.

\begin{lemma}\label{lm:ZeroModes}
\begin{equation*}
i\cL_\mu: \cT_{\eta_\mu}\cM_s \rightarrow \cT_{\eta_\mu}\cM_s
\end{equation*}
with $(i\cL_\mu)^2 X = 0$ for any vector $X\in \cT_{\eta_\mu}\cM_s.$
\end{lemma}

{\it Proof}.
Differentiating ${\mathcal E}_\mu' (T_a^{tr} \eta_\mu)=0$ with respect to $a$ and setting $a$ to zero gives 
\begin{equation}
\label{eq:ZeroModes1}
\cE''(\eta_\mu)\nabla_a\eta_\mu(x-a) |_{a=0}= \cL_\mu E_t = 0.  
\end{equation}
Similarly, differentiating ${\mathcal E}_\mu'(T_\gamma^g \eta_\mu)=0$ with respect to $\gamma$ and setting $\gamma$ to zero gives
\begin{equation}
\cL_\mu E_g = 0.
\end{equation}
Using (\ref{eq:NLEV}), we have 
\begin{equation}
\cL_\mu E_b = (-\Delta +\mu -f'(\eta_\mu)) ix \eta_\mu = -i\nabla_x \eta_\mu = iE_t,
\end{equation}
Furthermore, differentiating (\ref{eq:NLEV}) with respect to $\mu$ gives 
\begin{equation*}
(-\Delta+\mu-f'(\eta_\mu))E_s + \eta_\mu = 0,
\end{equation*}  
and hence
\begin{equation}
\label{eq:ZeroModes2}
\cL_\mu E_s = i(i\eta_\mu) = iE_g. \Box
\end{equation} 


\subsection{Two-soliton manifold}
We now discuss the manifold corresponding to two solitons. It is given by 
\begin{equation*}
\widetilde{\cM}_s^2:= \{(\eta_{\sigma_1}, \eta_{\sigma_2}), \; \sigma_i= (a_i,v_i,\gamma_i,\mu_i) \in \bbR^N\times\bbR^N\times [0,2\pi)\times I, \; i=1,2 \}.
\end{equation*}
The tangent space to $\widetilde{\cM}_s^2$ is 
\begin{equation*}
\cT_{(\eta_{\sigma_1},\eta_{\sigma_2})} \widetilde{\cM}^2_s = \{(X_1,X_2), \; X_i \in \cT_{\eta_{\sigma_i}}\cM_s, \; i=1,2\}.
\end{equation*}
We introduce the embedding mapping 
\begin{equation*}
{\mathsf E}: \; \widetilde{\cM}_s^2 \rightarrow H^1,
\end{equation*}
whose action on $\widetilde{\cM}_s^2$ and $\cT\widetilde{\cM}_s^2$ is given, respectively, by 
\begin{align*}
& {\mathsf E}(\eta_{\sigma_1},\eta_{\sigma_2}) = \eta_{\sigma_1} + \eta_{\sigma_2}\in H^1,\\
& {\mathsf E}(X_1,X_2) = X_1 + X_2 \in \cT H^1 + \cT H^1.
\end{align*}
In what follows, $\cM^2_s$ and $\cT\cM_s^2$ denote ${\mathsf E}(\widetilde{\cM}_s^2)$ and ${\mathsf E}(\cT\widetilde{\cM}_s^2)$ respectively.


\section{Skew-orthogonal decomposition}\label{sec:SOD}

Let $I$ be the same as in assumption (A3). We define 
\begin{equation*}
\Sigma := \{\sigma = (a,v,\gamma ,\mu) \in \bbR^N \times \bbR^N \times [0,2\pi) \times I \},
\end{equation*} 
and let 
\begin{equation*}
\Sigma^0 := \{\sigma = (a,v,\gamma ,\mu) \in \bbR^N \times \bbR^N \times [0,2\pi) \times I_0, \; \mathrm{with}\; \overline{I}_0 \subset I\backslash \partial I\; \mathrm{bounded} \}.
\end{equation*}
We define 
\begin{equation*}
\Sigma_{d,\kappa}^2 := \{(\sigma_1,\sigma_2)\in \Sigma^0\times \Sigma^0, \, \|a_1-a_2\|>d \ \ \mathrm{or} \ \ \|v_1-v_2\|>\kappa\}.
\end{equation*}
In other words, for $(\sigma_1,\sigma_2)\in \Sigma^2_{d,\kappa},$ the centers of mass of $\eta_{\sigma_1}$ and $\eta_{\sigma_2}$ are either separated by a distance larger than $d$ or their relative speed is larger than $\kappa.$

We consider the neighbourhood $U_{\delta, d, \kappa}\subset H^1$ defined by 
\begin{equation*}
U_{\delta, d, \kappa} := \{\psi\in H^1, \, \sup_{(\sigma_1,\sigma_2)\in \Sigma^2_{d,\kappa}}\|\psi - \eta_{\sigma_1} -\eta_{\sigma_2}\|_{L^2} < \delta \}.
\end{equation*}
We have the following proposition.

\begin{proposition}\label{pr:SOD}
Suppose (A2) and (A3) hold. Then, for $\delta\ll \inf_{\mu\in I_0}m'(\mu)$  and $\kappa \gg \frac{1}{\inf_{\mu\in I_0} m'(\mu)}$ (or $d \gg \max(\frac{1}{\sqrt{\mu_1}},\frac{1}{\sqrt{\mu_2}},|\log \inf_{\mu\in I_0} m'(\mu)|)$), there exist unique 
$$\sigma_1(\cdot),\sigma_2(\cdot) : \, U_{\delta,d,\kappa}\rightarrow \Sigma$$
such that 
\begin{equation}
\label{eq:sod1}
\psi = \eta_{\sigma_1(\psi)}+ \eta_{\sigma_2(\psi)} + w,
\end{equation}
and 
\begin{equation}
\label{eq:sod2}
\omega(w, X_i) = 0, \; i=1,2,
\end{equation}
for all $X_i \in \cT_{\eta_{\sigma_i}(\psi)}\cM_s,$ $i=1,2.$ 
\end{proposition}

{\it Proof}.
We define the mapping


$$G: \,U_{\delta, d,\kappa}\times \Sigma_{d,\kappa}^2 \rightarrow \bbR^{4N+4}$$
by 
\begin{equation*}
 G  (\psi, (\sigma_1,\sigma_2)):= 
\left (
\begin{matrix}
\omega(\psi - \eta_{\sigma_1} -\eta_{\sigma_2}, e_1 \eta_{\sigma_1}) \\
\cdot\\
\cdot\\
\cdot\\
\omega(\psi - \eta_{\sigma_1} -\eta_{\sigma_2}, e_{2N+2} \eta_{\sigma_1})\\
\omega(\psi - \eta_{\sigma_1} -\eta_{\sigma_2}, e_1 \eta_{\sigma_2}) \\
\cdot\\
\cdot\\
\cdot\\
\omega(\psi - \eta_{\sigma_1} -\eta_{\sigma_2}, e_{2N+2} \eta_{\sigma_2})
\end{matrix}
\right ).
\end{equation*} 
Then (\ref{eq:sod1}) and (\ref{eq:sod2}) are equivalent to $(\sigma_1(\psi),\sigma_2(\psi))$ satisfying, for a given $\psi\in U_{\delta ,d,\kappa},$ the equation
\begin{equation*}
G(\psi, (\sigma_1(\psi),\sigma_2(\psi))) = 0.
\end{equation*}

We use the implicit function theorem to show that there exist unique $\sigma_1(\psi),\sigma_2(\psi)\in \Sigma$ such that $ G  (\psi, (\sigma_1(\psi),\sigma_2(\psi)))=0.$

First, note that, by construction,
\begin{equation}
\label{eq:IFT1}
 G  (\eta_{\sigma_1}+\eta_{\sigma_2}, (\sigma_1,\sigma_2))=0.
\end{equation}
Furthermore, 
\begin{equation}
\label{eq:IFT2}
 G  \in C^1( U_{\delta, d,\kappa}\times \Sigma_{d,\kappa}^2 ; \bbR^{4N+4}),
\end{equation}
since it is linear in $\psi$ and $\eta_{\sigma_i},$ $i=1,2,$ and it is  differentiable in $\sigma_i\in \Sigma^0, \, i=1,2.$ We still need to show that $\partial_{(\sigma_1,\sigma_2)} G  (\eta_{\sigma_1}+\eta_{\sigma_2}, (\widetilde{\sigma}_1, \widetilde{\sigma_2}))|_{\widetilde{\sigma}_1 = \sigma_1 , \widetilde{\sigma}_2 = \sigma_2}$ is invertible for $\kappa \gg \frac{1}{\inf_{\mu\in I_0} m'(\mu)}$ (or $d \gg \max(\frac{1}{\sqrt{\mu_1}},\frac{1}{\sqrt{\mu_2}},\frac{1}{\inf_{\mu\in I_0} m'(\mu)})$).

We know that the matrix 
$$\{\omega(e_  \alpha   \eta_\sigma, e_\beta\eta_\sigma)\}_{  \alpha  ,\beta = 1}^{2N+2},$$ 
is invertible, see (\ref{eq:Metric}), Lemma \ref{lm:Metric}. 

We write
\begin{equation}
\label{eq:FastSlowDecomposition}
e_  \alpha  \eta_{\sigma_1}\overline{i e_\beta\eta_{\sigma_2}} =: e^{\frac{i}{2}(v_1-v_2)\cdot x} h_{  \alpha  \beta}(x),\,   \alpha  ,\beta=1,\cdots, 2N+2,
\end{equation}
which corresponds to a decomposition where the fast oscillating term (in space) $e^{\frac{i}{2}(v_1-v_2)\cdot x}$ is separated from the slowly oscillating term (in space) $h_{\alpha\beta}.$
Let $\|v_m\|:= \max(\|v_1\|,\|v_2\|,1).$ It follows from the fact that $f\in C^2$ (assumption (A2)) and the exponential localization in space of the solitons (assumption (A3)), that there exists $\xi\in (0,\min(\sqrt{\mu_1},\sqrt{\mu_2})),$ which is independent of $\|v_1-v_2\|,$  and a constant $C$ that dependends only on $\mu_1$ and $\mu_2,$ such that 
\begin{equation}
\label{eq:LocalDecayPrime}
\|h_{  \alpha  \beta}\|_{W^{3,1}(\bbR^N)} < C \|v_m\|^2 e^{-\xi\|a_1-a_2\|},
\end{equation}
for $\alpha  ,\beta=1,\cdots, 2N+2.$\footnote{More generally, if $f \in C^r(H^1,H^{-1}),$ 
$$\| h_{  \alpha  \beta}\|_{W^{r+1,1}(\bbR^N)} < C \|v_m\|^2 e^{-\xi\|a_1-a_2\|}.$$ For example, in the case of local nonlinearities, the above estimate holds for any $r\ge 1,$ in which case we obtain better estimates. }
Suppose that $\kappa\gg 1.$  Let $v:= v_1-v_2.$ Using that 
$$L e^{\frac{i}{2}v\cdot x} = e^{\frac{i}{2}v\cdot x},$$ where 
$$L:= -2i \frac{v}{\|v\|^2}\cdot\nabla_{x},$$ and integrating by parts three times, we obtain 
\begin{align}
\omega(e_  \alpha  \eta_{\sigma_1}, e_\beta\eta_{\sigma_2}) &= \int (L^3 e^{\frac{i}{2}v\cdot x}) h_{\alpha\beta}(x)~dx\nonumber \\
&= \int e^{\frac{i}{2}v\cdot x} (L^*)^3 h_{\alpha\beta}(x)~dx.
\label{eq:OmegaMixed}
\end{align}
Moreover,  
\begin{equation}
\|(L^*)^3  h_{\alpha\beta}\|_{L^1} \le \|v\|^{-3} \|h_{\alpha\beta}\|_{W^{3,1}}.
\label{eq:Mixed}
\end{equation}
Eqs. (\ref{eq:LocalDecayPrime}) - (\ref{eq:Mixed}) yield
\begin{align}
|\omega(e_  \alpha  \eta_{\sigma_1}, e_\beta\eta_{\sigma_2})|&\le C\|v\|^2\|v\|^{-3} \nonumber \\&\le C \|v\|^{-1} \nonumber \\ &\le C \kappa^{-1}. \label{eq:MixedEst}
\end{align}

 
(Suppose alternatively that $d\gg  \max(\frac{1}{\sqrt{\mu_1}},\frac{1}{\sqrt{\mu_2}})$ with $\|v_0\|=O(1)$ fixed. Then it follows from (\ref{eq:LocalDecayPrime}) that
$$|\omega(e_  \alpha  \eta_{\sigma_1}, e_\beta\eta_{\sigma_2})| < C e^{-\xi d},$$
for some positive constant $C$ that depends on $\mu_1$ and $\mu_2$ and $\xi\in (0,\min(\sqrt{\mu_1},\sqrt{\mu_2})).$) 

Hence, for $$\delta\ll \inf_{\mu\in I_0}m'(\mu)$$ and $$\kappa \gg (\inf_{\mu\in I_0}m'(\mu))^{-1}$$   ({\it or} $d\gg \max(\frac{1}{\sqrt{\mu_1}},\frac{1}{\sqrt{\mu_2}},|\log\inf_{\mu\in I_0}m'(\mu)|)$), the $(4N+4)\times(4N+4)$ matrix   
$$\partial_{(\sigma_1,\sigma_2)} G  (\eta_{\sigma_1}+\eta_{\sigma_2}, (\widetilde{\sigma}_1, \widetilde{\sigma_2}))|_{\widetilde{\sigma}_1 = \sigma_1 , \widetilde{\sigma}_2 = \sigma_2} = \left (
\begin{matrix}
\{\omega(e_  \alpha   \eta_{\sigma_1}, e_\mu\eta_{\sigma_1})\} & \{\omega(e_  \alpha   \eta_{\sigma_1}, e_\nu\eta_{\sigma_2})\} \\
\{\omega(e_  \alpha   \eta_{\sigma_2}, e_\mu\eta_{\sigma_1})\} & \{\omega(e_  \alpha   \eta_{\sigma_2}, e_\nu\eta_{\sigma_2})\} 
\end{matrix}
\right )$$ 
is invertible.

Invertibility of $\partial_{(\sigma_1,\sigma_2)} G  ,$ together with (\ref{eq:IFT1}), (\ref{eq:IFT2}) and the implicit function theorem, \footnote{See for example \cite{Mun}.} imply that there exist unique $C^1$ maps $\sigma_1(\psi)$ and $\sigma_2(\psi)$ such that $$ G  (\psi, (\sigma_1(\psi), \sigma_2(\psi)))=0. $$ $\Box$

\begin{remark}\label{rem:SODPrime} 
The group element $T_{av\gamma}\in {\mathsf H}^{2N+1}$ is given by 
\begin{equation*}
T_{av\gamma} = e^{-a\cdot \partial_x} e^{i\frac{v\cdot x}{2}} e^{i\gamma} .
\end{equation*}
It follows from (\ref{eq:CommutationRelations}) that $T^{-1}_{av\gamma} YT_{av\gamma} \in {\mathsf g}$ if $Y\in {\mathsf g}.$ Furthermore, it follows from translational invariance that $\omega(T_{av\gamma} u, T_{av\gamma} v)=\omega(u,v),$ for $u,v\in L^2.$ Therefore, we have from  Proposition \ref{pr:SOD} that  
\begin{equation}
\label{eq:SO}
\omega(w, Y (\eta_{\sigma_1}+ \eta_{\sigma_2})) = \omega ( w', Y' (\eta_{\sigma_1'}+ \eta_{\sigma_2'}) ) = 0, \nonumber 
\end{equation}
$\forall Y\in {\mathsf g},$ where $Y'=T^{-1}_{av\gamma} Y T_{av\gamma}\in {\mathsf g},$ $w'= T^{-1}_{av\gamma}w,$ and $\eta_{\sigma'}=T^{-1}_{av\gamma}\eta_\sigma.$
\end{remark}



\section{Reparametrized equations of motion}\label{sec:RepEqMot}

In this section, we apply the skew-orthogonal property to obtain reparametrized equations of motion for the parameters that characterize the projection of the true solution of (\ref{eq:NLSE}) with initial condition $\phi$ onto $\cM_s^2.$

We assume that the hypotheses for the skew-orthogonal decomposition, Sect. \ref{sec:SOD}, hold. We will verify in the proof of the main theorem that for large enough $\|v_0\|$ and small $h,$ this is indeed the case over a certain time interval.  

\begin{proposition}\label{pr:RepEqMot}
Consider (\ref{eq:NLSE}) with initial condition (\ref{eq:InitialCond}), and suppose that (A1)-(A6) hold. Assume further that there exists $\tau>0$ such that, for $t\in [0,\tau),$ $\psi(t),$ the solution of (\ref{eq:NLSE}) with initial condition $\phi,$ is in $U_{\delta,d,\kappa},$  where $\delta$ is given in Proposition \ref{pr:SOD}. Then, for $\|v_0\|\gg 1,$ there exists a positive constant $C_0<1$ independent of $\|v_0\|$ and $h,$ such that, for $\|w\|_{L^2}<C_0,$ the parameters $\sigma_i=(a_i,v_i,\gamma_i,\mu_i), \, i=1,2,$ satisfy the equations
\begin{align}
&\partial_t a_i = v_i + O(\|w\|^2_{L^2}+h^2+e^{-\xi\|a_1-a_2\|}),\label{eq:a} \\
&\partial_t v_i = - 2\nabla_{a_i} V_h(a_i,t)   + O(\|w\|^2_{L^2}+h^2+e^{-\xi\|a_1-a_2\|}) \label{eq:v} \\
&\partial_t \gamma_i = \mu_i + \frac{v_i^2}{4} - V_h(a_i,t)  + O(\|w\|^2_{L^2}+h^2+e^{-\xi\|a_1-a_2\|}),\\
&\partial_t \mu_i =   O(\|w\|^2_{L^2}+h^2+e^{-\xi\|a_1-a_2\|}), \label{eq:mu}
\end{align}
for some constant $\xi\in (0,\min(\sqrt{\mu_1},\sqrt{\mu_2}))$ that is independent of $\|v_0\|$ and $h.$
\end{proposition}

In what follows, we denote by $C$ a positive constant that is independent of $\|v\|$ and $h,$ but that may change from one line to another. 


{\it Proof}.
We first find the equation of motion for $$u_1=T^{-1}_{a_1v_1\gamma_1}\psi = e^{-\frac{i}{2}v_1\cdot x -i \gamma_1} \psi(x+a_1).$$ Using Proposition \ref{pr:SOD}, we have 
\begin{equation}
\label{eq:CenterMass1}
u_1 = \eta_{\mu_1} + \eta_{\sigma_2'} + w',
\end{equation}
where $\eta_{\sigma_2'}= T^{-1}_{a_1v_1\gamma_1} \eta_{\sigma_2}$ and $w'=T^{-1}_{a_1v_1\gamma_1} w.$ Here, $a_2' = a_2-a_1,$ $v_2' = v_2-v_1,$ $\gamma_2'=\gamma_2-\gamma_1,$ and $\mu_2'=\mu_2.$
It follows from Remark \ref{rem:SODPrime} that 
\begin{equation}
\label{eq:SODPrime}
\omega(w', X_1 + X_2) = 0,
\end{equation}
for all $X_1 \in \cT_{\eta_{\mu_1}}\cM_s$ and $X_2 \in \cT_{\eta_{\sigma_2'}}\cM_s.$

We define the coefficients 
\begin{align}
&c_j  : = \partial_t a_{1,j} - v_{1,j} , \; j=1,\cdots ,N , \nonumber\\
&c_{N+j}:=-\frac{1}{2} \partial_t v_{1,j}  - \nabla_{a_1} V_{h}(a_1,t)   , \; j=1,\cdots ,N ,\nonumber\\
&c_{2N+1} := \mu_1 -\frac{1}{4}v_1^2 + \frac{1}{2}\partial_t a_1 \cdot v_1 -   V_h(a_1,t) -\partial_t\gamma ,\nonumber \\
&c_{2N+2}:=-\partial_t \mu .\label{eq:Coefficients}
\end{align}

Note that
\begin{align}
e^{-\frac{i}{2}(v_1\cdot x +2\gamma_1)}\Delta \psi(x+a_1) &= \Delta u_1 + i v_1\cdot \nabla u_1 -\frac{v_1^2}{4} \\
e^{-\frac{i}{2} (v_1\cdot x + 2\gamma_1)}f(\psi(x+a_1)) &= f(u_1).\label{eq:EqMov}
\end{align}
Differentiating $u_1$ with respect to $t$ and using (\ref{eq:NLSE}), (\ref{eq:Coefficients})-(\ref{eq:EqMov}), we get
\begin{equation}
\partial_t u_1 = -i ((-\Delta + \mu_1)u_1 - f(u_1)) + \sum_{  \alpha  =1}^{2N+1} c_\alpha   e_\alpha   u_1 -i {\mathcal R}_V u_1, 
\end{equation}
where 
$${\mathcal R}_V = V_h(x+a_1,t) - V_h(a_1,t)- x\cdot \nabla V_h(a_1,t) .$$
In other words, 
\begin{equation}
\label{eq:CenterMassEq1}
\partial_t u_1 = -i \cE_{\mu_1}'(u_1) + \sum_{  \alpha  =1}^{2N+1} c_\alpha   e_  \alpha   u_1 -i\cR_V u_1,
\end{equation}
where $\cE_\mu$ is defined in (\ref{eq:EnergyFunctional}). 
Recall that 
\begin{equation*}
\cE_{\mu_1}'(\eta_{\mu_1}) = 0,
\end{equation*}
which implies 
\begin{equation}
\label{eq:CenterMassMin}
\cE_{\mu_1}' (u_1)= \cL_{\mu_1} (\eta_{\sigma_2'}+w') + N_{\mu_1} (\eta_{\sigma_2'}+w'), 
\end{equation}
where $$\cL_{\mu_1} = (-\Delta +\mu_1 -f'(\eta_{\mu_1})) \equiv \cE_{\mu_1}''(\eta_{\mu_1})$$ and $$N_{\mu_1} (\eta_{\sigma_2'}+w')= f(\eta_{\mu_1}+\eta_{\sigma_2'}+w') - f(\eta_{\mu_1}) -f'(\eta_{\mu_1})(\eta_{\sigma_2'}+w').$$
Substituting (\ref{eq:CenterMass1}) and (\ref{eq:CenterMassMin}) into (\ref{eq:CenterMassEq1}), we obtain
\begin{align}
\partial_t w' = &(-i\cL_{\mu_1} + \sum_{  \alpha  =1}^{2N+1}c_\alpha   e_  \alpha   -i \cR_V )w' + N_{\mu_1} (\eta_{\sigma_2'}+w') + \sum_{  \alpha  =1}^{2N+2} c_  \alpha   e_  \alpha   \eta_{\mu_1} -i \cR_V \eta_{\mu_1}  \nonumber \\ &-\partial_t\eta_{\sigma_2'}+( -i\cL_{\mu_1} + \sum_{  \alpha  =1}^{2N+1}c_  \alpha   e_  \alpha   -i \cR_V )\eta_{\sigma_2'} .\label{eq:CMEq}
\end{align}
To obtain the equations of motion for $a_1,v_1,\gamma_1$ and $\mu_1,$ we use the skew-orthogonal property to project (\ref{eq:CMEq}) onto $\cT_{\eta_{\mu_1}}\cM_s.$

It follows from (\ref{eq:SODPrime}) that $\langle iw', X \rangle = 0$ for all $X\in \cT_{\eta_{\mu_1}} \cM_s.$ Therefore,
\begin{equation}
\label{eq:SODT}
\partial_t \langle iw' , X\rangle = \partial_t \mu_1 \langle iw', \partial_{\mu_1} X\rangle + \langle i \partial_t w', X \rangle = 0.
\end{equation}  
Substituting the expression for $\partial_t w'$ given by (\ref{eq:CMEq}) in (\ref{eq:SODT}), and using  
\begin{equation}
\label{eq:SAGen}
e_  \alpha  ^* = -e_  \alpha  , \, \alpha=1,\cdots, 2N+2,
\end{equation}
we have
\begin{align}
\langle \cL_{\mu_1} w', X\rangle + &\langle i \sum_{  \alpha  =1}^{2N+2}c_  \alpha   e_  \alpha   w',X\rangle + \langle   \cR_V w',X\rangle + \langle iN_{\mu_1} (\eta_{\sigma_2'}+w'), X\rangle + \langle i\sum_{  \alpha  =1}^{2N+2} c_  \alpha   e_  \alpha   \eta_{\mu_1}, X\rangle +  \nonumber\\ &+\langle \cR_V \eta_{\mu_1} , X\rangle - \langle i \partial_t\eta_{\sigma_2'}, X\rangle + \langle ( \cL_{\mu_1} + i\sum_{  \alpha  =1}^{2N+1}c_  \alpha   e_  \alpha   + \cR_V )\eta_{\sigma_2'}, X\rangle = 0.\label{eq:Proj1}
\end{align}
Some of the terms in the above equation drop-out due to the zero modes of the Hessian. It follows from (\ref{eq:ZeroModes1})-(\ref{eq:ZeroModes2}), Lemma \ref{lm:ZeroModes}, that 
\begin{equation*}
X'= i\cL_{\mu_1}X \in \cT_{\eta_{\mu_1}\cM_s} \; \mathrm{if} \; X\in \cT_{\eta_{\mu_1}\cM_s},
\end{equation*}
and hence 
\begin{equation*}
\langle \cL_{\mu_1} w', X\rangle = \langle w', \cL_{\mu_1}X \rangle = -\omega(w,X')=0.
\end{equation*}
Together with (\ref{eq:SAGen}) and (\ref{eq:Proj1}), this yields 
\begin{align}
\sum_{  \alpha  =1}^{2N+2} c_  \alpha   \, \omega(e_  \alpha   \eta_{\mu_1}, X) = &\langle \cR_V \eta_{\mu_1} , X\rangle + \sum_{  \alpha  =1}^{2N+2}c_  \alpha    \, \langle i  e_  \alpha   w',X\rangle +\langle   \cR_V w',X\rangle + \langle iN_{\mu_1} (\eta_{\sigma_2'}+w'), X\rangle \nonumber \\& +\langle\cR_V \eta_{\sigma_2'}, X\rangle + \langle i\partial_t\eta_{\sigma_2'}, X\rangle +\langle \eta_{\sigma_2'},( \cL_{\mu_1} + i\sum_{  \alpha  =1}^{2N+1}c_  \alpha   e_  \alpha  ) X\rangle\label{eq:Proj2}.
\end{align}

We now estimate each term appearing in the right-hand-side of (\ref{eq:Proj2}) with $X = e_\beta   \eta_{\mu_1}, \, \beta=1,\cdots ,2N+2.$ Note that it follows from assumptions (A3) and (A6) that 
\begin{equation*}
\|\cR_Ve_\beta    \eta_{\mu_1}\|_{L^2} = O(h^2) , \; \beta =1,\cdots,2N+2,
\end{equation*}
and from (A3) that
\begin{align*}
&\|X\|_{L^2} = \|e_\beta   \eta_{\mu_1}\|_{L^2} = O(1),\; \beta=1,\cdots,2N+2\\
&\|e_  \alpha   X\|_{L^2} = \|e_\alpha e_\beta \eta_{\mu_1}\|_{L^2}= O(1), \, \alpha,\beta=1,\cdots, 2N+2.
\end{align*}
Hence, H\"older's inequality, (A3), (A6) and the fact that $V$ is real yield the estimates
\begin{align}
& |\langle \cR_V \eta_{\mu_1} , X\rangle| =  |\langle  \eta_{\mu_1} , \cR_V X\rangle| \le \|\eta_{\mu_1}\|_{L^2}\|\cR_Ve_\beta    \eta_{\mu_1}\|_{L^2}\le C h^2 \label{eq:EstLHS1}\\
&|\langle   \cR_V w',X\rangle|=|\langle   w',  \cR_V X\rangle|\le \|\cR_Ve_\beta    \eta_{\mu_1}\|_{L^2}\|w'\|_{L^2} \le C h^2 \|w'\|_{L^2}\\
&|\langle\cR_V \eta_{\sigma_2'}, X\rangle|=|\langle\eta_{\sigma_2'},\cR_V  X\rangle|\le \|\eta_{\sigma_2'}\|_{L^2}\|\cR_Ve_\beta    \eta_{\mu_1}\|_{L^2} \le Ch^2.
\end{align}
We also have from (A3) and H\"older's inequality that
\begin{align}
|\sum_{  \alpha  =1}^{2N+2}c_  \alpha    \, \langle i  e_  \alpha  w',X\rangle| &= |\sum_{  \alpha  =1}^{2N+2}  \,c_  \alpha   \langle i  w', e_  \alpha  X\rangle| \nonumber \\
& \le C \|c\| \|w'\|_{L^2}
\end{align}
where $\|c\|:=\max_{\alpha=1,\cdots ,2N+2}|c_\alpha|.$

We now use assumptions (A3)-(A4) to evaluate $|\langle iN_{\mu_1} (\eta_{\sigma_2'}+w'), X\rangle|.$  It follows from (A3) that $iX=ie_\beta \eta_{\mu_1}\in L^p, \, p\ge 1,$ which, together with (A5), yield     
\begin{equation*}
|\langle i (f(\eta_{\mu_1}+\eta_{\sigma_2'}+ w')- f(\eta_{\mu_1} + \eta_{\sigma_2'}) - f'(\eta_{\mu_1}+\eta_{\sigma_2'})w'), X\rangle| \le C\|w'\|_{L^2}^2.
\end{equation*}
It follows from the boundedness and the exponential localization of the solitons in space, (A3), and the fact that $f\in C^2,$ (A2), that 
\begin{align*}
\|\langle -i f'(\eta_{\mu_1})\eta_{\sigma_2'}, X\rangle| & \le\|f'(\eta_{\mu_1})\|_{L^\infty}\|\eta_{\sigma_2'}X\|_{L^1}\\&\le C e^{-\xi\|a_1-a_2\|},
\end{align*}
for some $\xi \in (0,\min(\sqrt{\mu_1},\sqrt{\mu_2}))$ which is independent of $\|v_0\|$ and $h.$
Moreover, it follows directly from (A4) that 
\begin{equation*}
|\langle i (f(\eta_{\mu_1} + \eta_{\sigma_2'})- f(\eta_{\mu_1})), X\rangle| \le C e^{-\xi\|a_1-a_2\|}
\end{equation*}
and
\begin{align*}
|\langle i (f'(\eta_{\mu_1}+ \eta_{\sigma_2'}) - f'(\eta_{\mu_1}) )w',X\rangle|  &\le \|w'\|_{L^2}\|(f'(\eta_{\mu_1}+ \eta_{\sigma_2'}) - f'(\eta_{\mu_1}))X\|_{L^2}\\&\le C e^{-\xi\|a_1-a_2\|}\|w'\|_{L^2}.
\end{align*}
Therefore,
\begin{equation}
|\langle iN_{\mu_1} (\eta_{\sigma_2'}+w'), X\rangle |\le C (\|w'\|_{L^2}^2 + e^{-\xi\|a_1-a_2\|}).
\end{equation}

To evaluate the remaining terms, we use the fact that $\eta_{\mu_1}$ and $\eta_{\sigma_2'}$ are exponentially localized in space, while their relative fast oscillating phase is $$\|v\|=\|v_1-v_2\|\ge \|v_m\|,$$ where $\|v_m\|:= \max(1,\|v_1\|,\|v_2\|).$  

When estimating an upper bound for $|\langle i \partial_t \eta_{\sigma_2'}, e_\beta    \eta_{\mu_1}\rangle|,$ the partial derivative with time contributes $\|v_m\|^2,$ since 
\begin{equation}
\label{eq:PartDerEta}
\partial_t \eta_{\sigma} = \left [ \sum_{  j  =1}^N \partial_t a_j   e_j  + \frac{1}{2}\sum_{j =N+1}^{2N} \partial_t  v_j   e_j   + (\partial_t\gamma + \frac{v\cdot\dot{a}}{2}) e_{2N+1} + \partial_t\mu e_{2N+2} 
\right ]\eta_\sigma.
\end{equation}
However, using (\ref{eq:FastSlowDecomposition}) and (\ref{eq:LocalDecayPrime}), and integrating by parts twice in space, we can pull a factor of $\|v_1-v_2\|^{-2}$ from the fast oscillating term $e^{\frac{i}{2}(v_2-v_1)\cdot x},$ see the discussion below (\ref{eq:LocalDecayPrime}) in the proof of Proposition \ref{pr:SOD}. Hence 
\begin{equation}
|\langle \partial_t \eta_{\sigma_2'}, e_\beta    \eta_{\mu_1}\rangle|\le C e^{-\xi\|a_1-a_2\|}.
\end{equation}
Furthermore, (A2) and (A3) yield
\begin{equation}
|\langle \eta_{\sigma_2'},\cL_{\mu_1} e_\beta \eta_{\mu_1}\rangle|\le C e^{-\xi\|a_1-a_2\|}.
\end{equation}
Again, using (\ref{eq:FastSlowDecomposition}) and (\ref{eq:LocalDecayPrime}) and integrating by parts twice in space to pull a factor of $\|v_1-v_2\|^{-2}$ from the fast oscillating factor $e^{\frac{i}{2}(v_2-v_1)\cdot x},$ we have
\begin{equation}
\label{eq:EstF}
|\langle \eta_{\sigma_2'},( i\sum_{  \alpha  =1}^{2N+1}c_  \alpha   e_  \alpha  ) e_\beta \eta_{\mu_1}\rangle|\le C \|c\|\|v\|^{-2} e^{-\xi\|a_1-a_2\|}.
\end{equation}
From (\ref{eq:Proj2}) - (\ref{eq:EstF}), we have 
\begin{equation}
|\sum_{  \alpha  =1}^{2N+2} c_  \alpha   \, \omega(e_  \alpha   \eta_{\mu_1}, e_\beta    \eta_{\mu_1})| \le  C[\|w\|_{L^2}^2 + \|c\| (\|w\|_{L^2}+\|v\|^{-2})+h^2+ e^{-\xi\|a_1-a_2\|}],\label{eq:ProjEst1}
\end{equation}
for $\beta=1,\cdots,2N+2,$ where we used  $$\|w'\|_{L^2}= \|T^{-1}_{a_1v_1\gamma_1}w\|_{L^2} = \|w\|_{L^2}$$ due to translational invariance.

Using Lemma \ref{lm:Metric}, (\ref{eq:Metric}) and (\ref{eq:ProjEst1}), and assuming $\|w\|_{L^2}$ and $\|v\|^{-2}\le \frac{1}{4C}\|\Omega_{\mu_1}\|,$  we obtain the estimate 
$$\|c\|\le C (\|w\|_{L^2}^2 +h^2+ e^{-\xi\|a_1-a_2\|}).$$
Recalling now the definition of $c_\alpha, \alpha =1,\cdots,2N+2$ (see (\ref{eq:Coefficients})), we conclude (\ref{eq:a}) - (\ref{eq:mu}), with  $i=1.$ 

To get the equations of motion for $a_2,v_2,\gamma_2$ and $\mu_2,$ we consider $u_2= T^{-1}_{a_2v_2\gamma_2}\psi,$ and we repeat the above analysis with $1\leftrightarrow 2.$ $\Box$


\section{Control of the fluctuation}\label{sec:ControlFluctuation}

We now control the $L^2$-norm of the fluctuation $w$ using conservation of charge, the skew-orthogonal property, Sect. \ref{sec:SOD}, and the reparametrized equations of motion, Sect. \ref{sec:RepEqMot}.   
\begin{proposition}\label{pr:ControlFluctuation}
Consider (\ref{eq:NLSE}) with initial condition (\ref{eq:InitialCond}), and suppose that (A1)-(A6) hold. Assume further that there exists $\tau>0$ such that, for $t\in [0,\tau),$ $\psi(t)\in U_{\delta,d,\kappa},$  where $\delta$ is given in Proposition \ref{pr:SOD}. Then, for $\|v_0\|\gg 1$ and $h\ll 1,$   
\begin{equation*}
\sup_{t\in [0,C  \alpha  \min(\log \|v_0\|, 2 |\log 
h|))}\|w(t)\|^2_{L^2} \le C'(\|v_0\|^{-1+  \alpha  }+h^{2(1-  \alpha  )}),
\end{equation*}
for some positive constants $C$ and $C'$ that are independent of $v_0, h,$ and $  \alpha  \in (0,1).$
\end{proposition}

{\it Proof}. From conservation of charge ($L^2$-norm) of the solution of (\ref{eq:NLSE}),
$$\|\psi(t)\|_{L^2}=\|\phi\|_{L^2},$$
and skew-orthogonal decomposition (Proposition \ref{pr:SOD}), we have
\begin{equation}
\label{eq:ChargeCons}
\|\psi\|_{L^2}^2 = \|w\|_{L^2}^2 + \|\eta_{\mu_1}\|_{L^2}^2 +\|\eta_{\mu_2}\|^2_{L^2} + 2\Re \langle \eta_{\sigma_1},\eta_{\sigma_2}\rangle = \|\phi\|_{L^2}^2 ,
\end{equation}
where we used
\begin{equation*}
\langle w, \eta_{\sigma_j}\rangle = -\omega (w,i\eta_{\sigma_j}) = 0, 
\end{equation*}
and
\begin{equation*}
\|\eta_{\sigma_j}\|_{L^2} = \|\eta_{\mu_j}\|_{L^2},
\end{equation*}
for $j=1,2.$ 

Differentiating (\ref{eq:ChargeCons}) with respect to $t,$ and recalling that $m(\mu)= \frac{1}{2}\|\eta_\mu\|_{L^2}^2,$
we get
\begin{equation}
\label{eq:PartEqFluc}
\partial_t\|w\|_{L^2}^2 = - 2\partial_t \mu_1 \, \partial_{\mu_1} m(\mu_1) - 2\partial_t \mu_2 \, \partial_{\mu_2}m(\mu_2) - 2\partial_t\Re \langle \eta_{\sigma_1},\eta_{\sigma_2}\rangle.
\end{equation}

First, using the exponential localization of solitons in space and the fast relative phase of the solitons, we estimate an upper bound for $$|\partial_t \Re\langle \eta_{\sigma_1},\eta_{\sigma_2}\rangle |=|\partial_t\omega(\eta_{\sigma_1}, e_{2N+1}\eta_{\sigma_2})| . $$ 
From (\ref{eq:a})-(\ref{eq:mu}) and (\ref{eq:PartDerEta}), it seems a priori that  $|\partial_t \Re\langle \eta_{\sigma_1},\eta_{\sigma_2}\rangle | $ is of order $\|v\|^2.$  However, we can pull a factor of $\|v_1-v_2\|^{-2}$ from the fast oscillating phase $e^{\frac{i}{2}(v_1-v_2)\cdot x}$ by integrating by parts twice, as in (\ref{eq:LocalDecayPrime}) - (\ref{eq:MixedEst}) in Sect. \ref{sec:SOD}. Therefore,
\begin{equation}
|\partial_t \Re\langle \eta_{\sigma_1},\eta_{\sigma_2}\rangle | \le C e^{-\xi\|a_1-a_2\|}.
\end{equation}
Furthermore, (\ref{eq:mu}) implies that 
\begin{equation}
|\partial_t \mu_1 \, \partial_{\mu_1} m(\mu_1) + \partial_t \mu_2 \, \partial_{\mu_2}m(\mu_2)|\le C (h^2 + e^{-\xi\|a_1-a_2\|}+\|w\|_{L^2}^2).
\label{eq:EstMu}
\end{equation}

Now, (\ref{eq:PartEqFluc}) - (\ref{eq:EstMu}) yield
\begin{equation}
\label{eq:PartEqFluc2}
|\partial_t \|w\|_{L^2}^2| \le C (h^2 + e^{-\xi\|a_1-a_2\|}+\|w\|_{L^2}^2),
\end{equation}
for some positive constant $C$ independent of $\|v_0\|$ and $h.$ 

It follows from (\ref{eq:PartEqFluc2}) and the Duhamel formula that 
\begin{equation}
\|w\|_{L^2}^2 \le C (e^{ct}(h^2+\|\widetilde{w} \|^2_{L^2}) + \int_0^t ds\, e^{c(t-s)} e^{-\xi\|a_1-a_2\|}).\label{eq:PartFluct21}
\end{equation} 
For times $t<C\|v_0\|^{\epsilon},$ $\epsilon\in (0,1),$ we know from (\ref{eq:v}) that 
\begin{equation}
\|v_1(t)-v_2(t)\|\ge c_0\|v_0\| ,\label{eq:RelSpeed}
\end{equation}
for some constant $c_0>0.$
Making the change of variables 
$$s\rightarrow a(s),$$
where 
$$a(s):=\|a_1(s)-a_2(s)\| ,$$ 
and using that 
$$e^{-\xi a(s)} = -\frac{1}{\xi \partial_s a(s)}\partial_s e^{-\xi a(s)},$$
(\ref{eq:a}), (\ref{eq:v}) and (\ref{eq:RelSpeed}), we have
$$|\int_0^t ds\, e^{c(t-s)} e^{-\xi\|a_1-a_2\|}| \le C\frac{e^{ct}}{\|v_0\|}.$$ Together with (\ref{eq:PartFluct21}), we get the estimate 
\begin{equation}
\|w\|_{L^2}^2 \le C (h^2 e^{ct} + \frac{1}{\|v_0\|} e^{ct}),\label{eq:PartEqFluc3}
\end{equation}
for some positive constants $C$ and $c$ that are independent of $\|v_0\|$ and $h.$ Let $\tau:= \frac{  \alpha  }{c}\min(\log\|v_0\|,2|\log h|)$ for some $  \alpha  \in (0,1).$  For $t< \tau,$ (\ref{eq:PartEqFluc3}) implies
\begin{equation*}
\sup_{t\in[0,\tau)}\|w\|_{L^2}^2 < C (\|v_0\|^{-1+  \alpha  } +h^{2(1-  \alpha  )}) .
\end{equation*}
$\Box$


\section{Proof of Theorem \ref{th:Main}}\label{sec:Proof}

We now show that, for $\|v_0\|\gg 1$ large enough and $h\ll 1$ small enough, the hypotheses of Propositions \ref{pr:SOD}, \ref{pr:RepEqMot} and \ref{pr:ControlFluctuation} can be simultaneously satisfied.

Let $$T:= \sup \{t\ge 0, \; \psi(t) \in U_{d,\kappa,\delta} \; {\mathrm with}\; \delta \; \mathrm{as}\; \mathrm{in}\; \mathrm{Proposition} \; \ref{pr:SOD}\}.$$ By continuity of $\|w(t)\|_{L^2},$ $T>0.$ If $T\le C\alpha  \min(\log\|v_0\|,2|\log h|),$ then by Proposition  \ref{pr:ControlFluctuation},
\begin{equation}
\label{eq:SupFluctuation2}
\sup_{t\in [0,T)} \|w (t)\|_{L^2} \le C' (\|v_0\|^{\frac{-1+  \alpha  }{2}} + h^{1-  \alpha  }).
\end{equation}
Here, $C,C'$ appear in Proposition \ref{pr:ControlFluctuation}. We need 
\begin{equation*}
\delta  \ll \inf_{\mu\in I_0}m'(\mu),
\end{equation*}
where $\delta$ appears in Proposition \ref{pr:SOD}.
Consider $v_0$ and $h$ satisfying
$$C'(\|v_0\|^{\frac{-1+\alpha}{2}}+ h^{1-\alpha}) \le \frac{\delta}{2} \ll (\inf_{\mu\in I_0} m'(\mu))^{\frac{2}{1-\alpha}} \ll  \inf_{\mu\in I_0} m'(\mu).$$
Then $\|w(T)\|_{L^2}\le \frac{\delta}{2},$ and $T$ is not the maximal time unless $T=C\alpha  \min(\log\|v_0\|,2|\log h|).$ Then (\ref{eq:SupFluctuation2}) yields (\ref{eq:SupFluctuation}). Furthermore, the hypotheses of Proposition \ref{pr:RepEqMot} are satisfied. Using (\ref{eq:SupFluctuation}) in (\ref{eq:a})-(\ref{eq:mu}) gives the estimates on the evolution of the parameters in Theorem \ref{th:Main}. $\Box$


\section{Comments on separating solitons}\label{rem:ProofLongerTime}
We now discuss Remark \ref{rem:LongerTime} in Subsect. \ref{sec:Main}, whose hypotheses we assume. 

\begin{itemize}
\item[(1)]Suppose that the soliton centers of mass are initially separated by a distance $d \gg \max(\frac{1}{\sqrt{\mu_1}},\frac{1}{\sqrt{\mu_2}},|\log \inf_{\mu\in I_0} m'(\mu)|),$ such that $\|\widetilde{w} \|_{L^2} = O(e^{-\chi d})$ for some $\chi>0,$ and that $\|v_1\|,\|v_2\| = O(1).$ Then the analysis above (Propositions \ref{pr:SOD}, \ref{pr:RepEqMot} and \ref{pr:ControlFluctuation}) holds, except that (\ref{eq:PartEqFluc2}) implies
$$\|w\|_{L^2}^2 \le C e^{-C'd + C''d^{\epsilon}} < Cd^{-1},$$ for $t< d^\epsilon,\, \epsilon\in(0,1),$ from which follows the claim of (1) of Remark \ref{rem:LongerTime}.

\item[(2) ] In the case of escaping solitons,  (\ref{eq:PartEqFluc2}) in the proof of Proposition \ref{pr:ControlFluctuation} implies that 
$$\|w\|_{L^2}^2 \le C e^{-C'\|v_0\| + C''\|v_0\|^{\epsilon}} < C/\|v_0\|$$ for $t\le \|v_0\|^{\epsilon},$ $\epsilon\in (0,1).$ Hence the claim (2) of Remark \ref{rem:LongerTime} also holds. 
\end{itemize}


\section{Appendix}

\noindent {\it Proof of Lemma \ref{lm:Metric}, Sect. \ref{sec:NLSeq}}.
Explicitly, $$\Omega_{\sigma}|_{\cT_{\eta_\sigma}\cM_s} := \{\langle e_  \alpha  \eta_\sigma, i e_\beta   \eta_\sigma \rangle\}_{1\le \alpha,\beta\le 2N+2},$$
where $e_\alpha   \eta_\sigma, \alpha=1,\cdots , 2N+2,$ are basis vectors of $\cT_{\eta_\sigma} \cM_s.$
For $\alpha=1,\cdots, N$ and $\beta=N+1,\cdots, 2N.$
\begin{align*}
\langle e_  \alpha  \eta_\sigma, i e_\beta    \eta_\sigma\rangle &= \langle -\partial_{ x_  \alpha  }( e^{\frac{i}{2}v\cdot (x-a)+i\gamma}\eta_{\mu}(x-a)), - x_\beta e^{\frac{i}{2}v\cdot (x-a)+i\gamma}\eta_{\mu}(x-a) \rangle \\ &= \frac{ v_  \alpha  }{2}\langle i\eta_\mu(x-a),  x_\beta \eta_\mu(x-a)\rangle + \langle \partial_{ x_  \alpha  }\eta_\mu(x-a),  x_\beta \eta_\mu(x-a)\rangle.   
\end{align*}
It follows from translational invariance of the integral, and positivity and spherical symmetry of $\eta_\mu(x)$ that 
\begin{equation*}
\langle i\eta_\mu (x-a),  x_\beta \eta_\mu (x-a)\rangle = \langle i \eta_\mu  (x),  x_\beta \eta_\mu (x)\rangle + \langle i\eta_\mu (x),  a_\beta   \eta_\mu (x)\rangle = 0 , 
\end{equation*}
and, by integration by parts, 
\begin{align*}
\langle \partial_{ x_  \alpha  }\eta_\mu(x-a),  x_\beta \eta_\mu(x-a)\rangle &=  -\delta_{\alpha\beta} \langle \eta_\mu(x), \eta_\mu(x)\rangle -\langle  x_\beta\eta_\mu(x-a), \partial_{ x_  \alpha  } \eta_\mu(x-a)\rangle\\ &= -2\delta_{\alpha\beta} m(\mu) - \langle \partial_{ x_  \alpha  }\eta_\mu(x-a),  x_\beta \eta_\mu(x-a)\rangle,
\end{align*}
where $m(\mu) = \frac{1}{2}\|\eta_\mu\|_{L^2}$ and $\delta_{\alpha\beta}$ stands for the Kroenecker delta. Therefore,
\begin{equation*}
\langle e_  \alpha   \eta_\mu, ie_\beta    \eta_\mu\rangle = - \langle e_\beta    \eta_\sigma, ie_  \alpha  \eta_\sigma\rangle = -\delta_{\alpha\beta} m(\mu), \, \alpha=1,\cdots, N, \, \beta=N+1, \cdots, 2N.
\end{equation*}
Furthermore, 
\begin{align*}
\langle \partial_\mu\eta_\sigma , i\partial_\gamma \eta_\sigma\rangle &= -\langle \partial_\mu \eta_\mu(x-a), \eta_\mu (x-a)\rangle \\
&= -\partial_\mu m(\mu),
\end{align*}
and hence 
\begin{equation*}
\langle e_{2N+2}\eta_\sigma, ie_{2N+1}\eta_\sigma\rangle = -\langle e_{2N+1}\eta_\sigma, ie_{2N+2}\eta_\sigma\rangle= - m'(\mu),
\end{equation*}
where $m'(\mu)=\partial_\mu m(\mu).$ 
For $\alpha,\beta = 1,\cdots, N,$
\begin{align*}
\langle e_  \alpha  \eta_\sigma, i e_\beta   \eta_\sigma \rangle &= \langle \partial_{ x_  \alpha  } (e^{\frac{i}{2}v\cdot (x-a)+i\gamma}\eta_\mu(x-a)), i \partial_{ x_\beta} (e^{\frac{i}{2}v\cdot (x-a)+i\gamma}\eta_\mu(x-a) )\rangle \\
&= \langle (\frac{i}{2}  v_  \alpha   + \partial_{ x_  \alpha  })\eta_\mu(x), i (\frac{i}{2}  v_\beta + \partial_{ x_\beta})\eta_\mu(x)\rangle \\
&= \frac{1}{4} v_  \alpha   v_\beta \langle \eta_\mu, i\eta_\mu\rangle + \langle \partial_{ x_  \alpha  } \eta_\mu, i \partial_{ x_\beta} \eta_\mu\rangle + \frac{ v_  \alpha  }{2}\langle \eta_\mu, \partial_{ x_\beta}\eta_\mu\rangle + \frac{ v_\beta}{2} \langle \partial_{ x_  \alpha  }\eta_\mu, \eta_\mu\rangle ,
\end{align*}
where we used translational invariance in the second line. It follows from spherical symmetry of $\eta_\mu(x)$ that 
\begin{align*}
&\langle \eta_\mu, \partial_{ x_\beta}\eta_\mu\rangle = 0,\\
&\langle \partial_{ x_  \alpha  } \eta_\mu, i \partial_{ x_\beta} \eta_\mu\rangle = 0, \; \alpha\ne \beta.
\end{align*}
Furthermore, since $\eta_\mu$ is real, 
$$\langle \eta_\mu, i\eta_\mu\rangle = \langle \partial_{ x_  \alpha  } \eta_\mu, i \partial_{ x_  \alpha  } \eta_\mu\rangle=0.$$ Therefore, 
\begin{equation*}
\langle e_  \alpha   \eta_\sigma, ie_\beta    \eta_\sigma \rangle = 0, \, \alpha,\beta = 1,\cdots,N.
\end{equation*}
For $\alpha=1,\cdots, N,$
\begin{equation*}
\langle e_  \alpha  \eta_\sigma, i e_{2N+1}\eta_\sigma\rangle = \langle (\frac{i}{2} v_  \alpha   + \partial_{ x_  \alpha  })\eta_{\mu}(x-a), \eta_{\mu}(x-a)\rangle = 0.
\end{equation*}
and
\begin{equation*}
\langle e_  \alpha  \eta_\sigma, i e_{2N+2}\eta_\sigma\rangle = -\langle (\frac{i}{2} v_  \alpha   + \partial_{ x_  \alpha  })\eta_{\mu}(x-a), i\partial_\mu \eta_{\mu}(x-a)\rangle = - \frac{1}{2}v_  \alpha   m'(\mu).
\end{equation*}
For $\alpha=N+1,\cdots, 2N,$
\begin{equation*}
\langle e_  \alpha  \eta_\sigma, i e_{2N+1}\eta_\sigma\rangle = -\langle i x_  \alpha  \eta_\mu(x-a), \eta_\mu(x-a)\rangle = 0,
\end{equation*}
and
\begin{equation*}
\langle e_  \alpha  \eta_\sigma, i e_{2N+2}\eta_\sigma\rangle = \langle i x_  \alpha  \eta_\mu(x-a), i\partial_\mu \eta_\mu(x-a)\rangle = a_  \alpha   m'(\mu). 
\end{equation*}
Explicitly, we have
\begin{equation*}
\Omega_{\sigma}|_{\cT_{\eta_{\sigma}}\cM_s} = 
\left(
\begin{matrix}
0 & -m(\mu) {\mathbf 1}_{N\times N} & 0 & -\frac{1}{2}vm'(\mu) \\
m(\mu) {\mathbf 1}_{N\times N} & 0 & 0 & am'(\mu) \\
0 & 0 & 0 & m'(\mu) \\
\frac{1}{2}v^T m'(\mu) & -a^Tm'(\mu) & -m'(\mu) & 0
\end{matrix}
\right),
\end{equation*}
where ${\mathbf 1}_{N\times N}$ is the $N\times N$ identity matrix, and $(\cdot)^T$ stands for the transpose of a vector in $\bbR^N.$ One may easily verify that the $(2N+2)\times (2N+2)$ skew-symmetric matrix $\Omega_\sigma$ given in (\ref{eq:Metric}) is invertible if $\partial_\mu m(\mu)>0.$$\Box$


\end{document}